# Behavioural Transformation to Improve Circuit Performance in High-Level Synthesis*

R. Ruiz-Sautua, M. C. Molina, J.M. Mendías, R. Hermida
Dpto. Arquitectura de Computadores y Automática
Universidad Complutense de Madrid
rsautua@fdi.ucm.es, {cmolinap, mendias, rhermida}@dacya.ucm.es

**Abstract**

*Early scheduling algorithms usually adjusted the clock cycle duration to the execution time of the slowest operation. This resulted in large slack times wasted in those cycles executing faster operations. To reduce the wasted times multi-cycle and chaining techniques have been employed. While these techniques have produced successful designs, its effectiveness is often limited due to the area increment that may derive from chaining, and the extra latencies that may derive from multicycling. In this paper we present an optimization method that solves the time-constrained scheduling problem by transforming behavioural specifications into new ones whose subsequent synthesis substantially improves circuit performance. Our proposal breaks up some of the specification operations, allowing their execution during several possibly unconsecutive cycles, and also the calculation of several data-dependent operation fragments in the same cycle. To do so, it takes into account the circuit latency and the execution time of every specification operation. The experimental results carried out show that circuits obtained from the optimized specification are on average 60% faster than those synthesized from the original specification, with only slight increments in the circuit area.*

## 1. Introduction

A High–Level Synthesis (HLS) process transforms the behavioural description of a circuit into a Register-Transfer-Level (RTL) implementation. It involves three major tasks: scheduling, allocation, and binding. Scheduling determines the number of clock cycles (latency) and their duration, and assigns operations of the behavioural description to them. Allocation selects a set of functional, storage, and routing resources from the components library. And binding assigns operations to functional units (FUs), variables to storage elements, and data transfers to routing resources.

Early algorithms used to propose schedules with at least as many cycles as the number of operations in the critical path. The clock cycle duration usually equals the longest arrival time of the result bits of the specification operations. This produces a large slack to be wasted in those cycles where the results calculated have smaller arrival times than the cycle length. Additionally, some datapath FUs remain idle during part of the clock cycle if the results calculated have different arrival times.

Many efforts in high-level scheduling have been concentrated on improving circuit performance (time required to execute all the behavioural description operations) by minimizing the slack times wasted in clock cycles. Traditionally, pipelining has been the preferred technique to improve system performance, although it does not reduce the circuit latency [1-2]. In order to reduce the latency, some algorithms have added some optimization phases after the scheduling process to adjust either the number or duration of the clock cycles [3-6]. The algorithms presented in [4] and [5] reduce the circuit latency by allocating respectively carry-save and variable-latency operators (the time taken to compute the outputs depends on the input values). In [6] the phase coupling problem of the HLS is alleviated by allowing the later adjustment of every scheduling decision.

Most scheduling algorithms have reduced circuit latency by incorporating chaining and multi-cycle features. Chaining helps to reduce the number of clock cycles by allowing the execution of several data-dependent operations in the same cycle. The result produced by one operation is supplied as input operand to another operation in the same cycle. This technique requires more FUs (the chained operations cannot share HW resources) and less storage units (the intermediate results are not stored). One step further, the bit-level chaining (BLC) [3] [7] exploits the inherent parallelism of data-dependent operations with rippling effect (e.g. additions and multiplications). Thus, part of these chained operations can be executed in parallel at the bit level. Multi-cycle reduces the clock cycle duration by allowing the execution of long operations across several consecutive cycles. In this case, the results produced need several cycles to be available. Non-integer multi-cycle has been used in [3] to chain the result produced in one cycle by a multi-cycle operator to the next data-dependent operation.

Although all these design techniques reduce the circuit latency, in most cases better results could be obtained if:

---

* This work has been supported by Grant CICYT TIC-2002/750

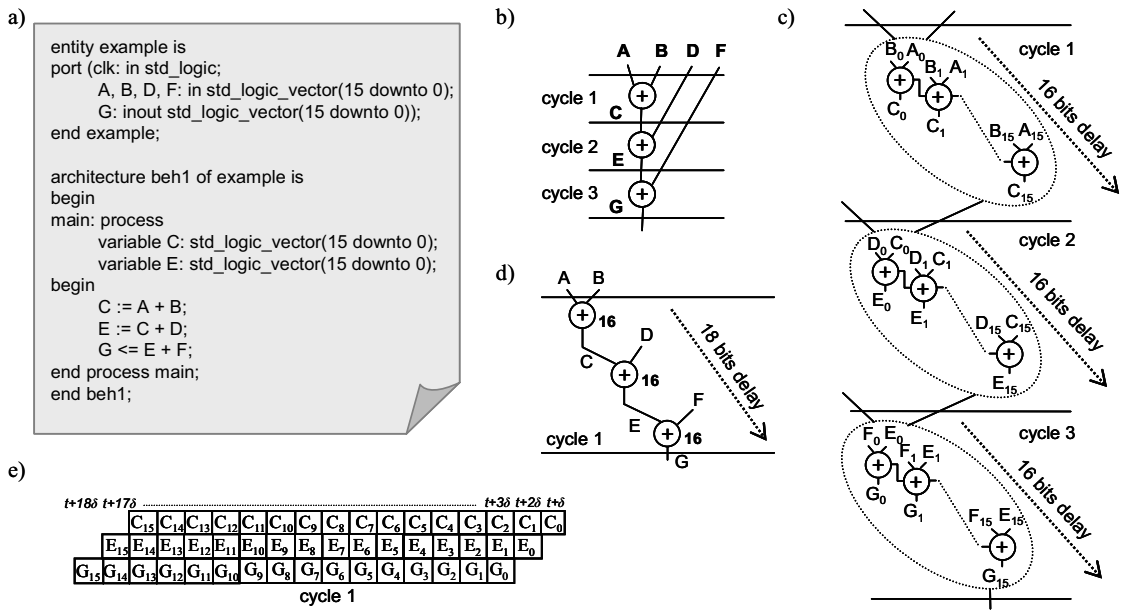

Fig. 1. a) Behavioural specification, b) conventional schedule, c) zoom of the conventional schedule, d) schedule using operation chaining, e) inherent parallelism of the chained operations that calculate *C*, *E*, and *G*.

- clock cycle duration were *independent* of the execution times of operations
- the execution of operations across several unconsecutive cycles were allowed
- every result bit were available (to be used as an input operand) the cycle it is calculated in, even if the overall execution of the operation has not finished

In this paper we present an optimization method that takes into account the above features. It substitutes, before synthesis, some of the specification operations for several ones whose types and widths may be different from the original, and that can be scheduled independently. The schedule of the new operations considerably reduces the slack times wasted, as compared with the implementations synthesized from the original specification. It produces implementations with the following features:

- one original operation may be executed in several unconsecutive cycles
- one operation may start its execution before the computation of its predecessors has been completed

## 2. Motivational example

Figures 1 and 2 illustrate an example of how this optimization method may improve circuit performance. A behavioural specification written in VHDL is shown in Fig. 1 a). It consists of 3 data-dependent additions of 16 bits. Fig. 1 b) presents the schedule proposed by a conventional algorithm, where every addition has been scheduled in a different clock cycle. Fig. 1 c) illustrates a zoom of this schedule. It clearly shows that the execution time of every 16-bits addition is equivalent to the time needed to execute 16 chained 1-bit additions. Hence, the execution time of all the specification operations is equivalent to the execution time of 48 chained 1-bit additions (16×3 cycles). The datapath synthesized from this schedule is formed by one 16-bits adder. It corresponds to the circuit with minimal FUs area, but maximal execution time.

Fig. 1 d) illustrates another possible schedule using BLC. In this case, the execution time is equivalent to the time required to execute 18 chained 1-bit additions, thanks to the rippling effect of additions that allows the execution in parallel of some bits of the 3 operations. Fig. 1 e) shows in every column the addition bits that are executed in parallel, and the time when every result bit is available in function of the delay $\delta$ of 1-bit adder (above each column). For example, bits $i$ of $C$, $i-1$ of $E$ and $i-2$ of $G$ are calculated simultaneously. If the execution starts in time $t$, then bit $i$ of $C$ is available in time $t+(i+1)\cdot\delta$. The datapath synthesized from this schedule consists of 3 chained adders of 16 bits. It corresponds to the circuit with minimum execution time, but maximal FUs area.

Fig. 2 a) shows our transformed specification. It has been obtained taking into account the circuit latency, and the number of bits of every addition that can be executed simultaneously. In the transformed specification every addition has been substituted for 3 data-dependent smaller additions with similar execution times. Fig. 2 b) shows the schedule obtained by a conventional algorithm from the new specification, where a fragment of every original addition has been scheduled in every cycle. Fig. 2 c) illustrates the addition bits that are calculated simultaneously in every cycle, being the execution time equivalent to 18 chained 1-bit adders (6×3 cycles). Note that the clock cycle duration achieved (equivalent to 6 chained 1-bit additions) is independent of the operation execution time (16 chained 1-bit additions). The datapath obtained from the allocation of this schedule comprises 3 chained adders of 6 bits, and every adder is dedicated to calculate just one addition in the behavioural description. For example, one adder calculates $C_{5..0}$ in the first cycle, $C_{11..6}$ in the second one, and $C_{15..12}$ in the third. Note that the storage area (5 registers of 1 bit) is quite smaller as well, because most result bits calculated in every cycle are also consumed in that same cycle to compute some result bits of another chained operation. The dedicated registers needed to stabilize the input and output ports have not been considered because they coincide in both implementations. For example, in the first cycle one adder calculates $C_{5..0}$, where $C_{4..0}$ is used as input operand by a second adder, which calculates $E_{4..0}$, and $E_{3..0}$ is used as input operand by the third adder. Therefore just $C_5$ and $E_4$ plus the 3 carry outs must be stored in this first clock cycle.

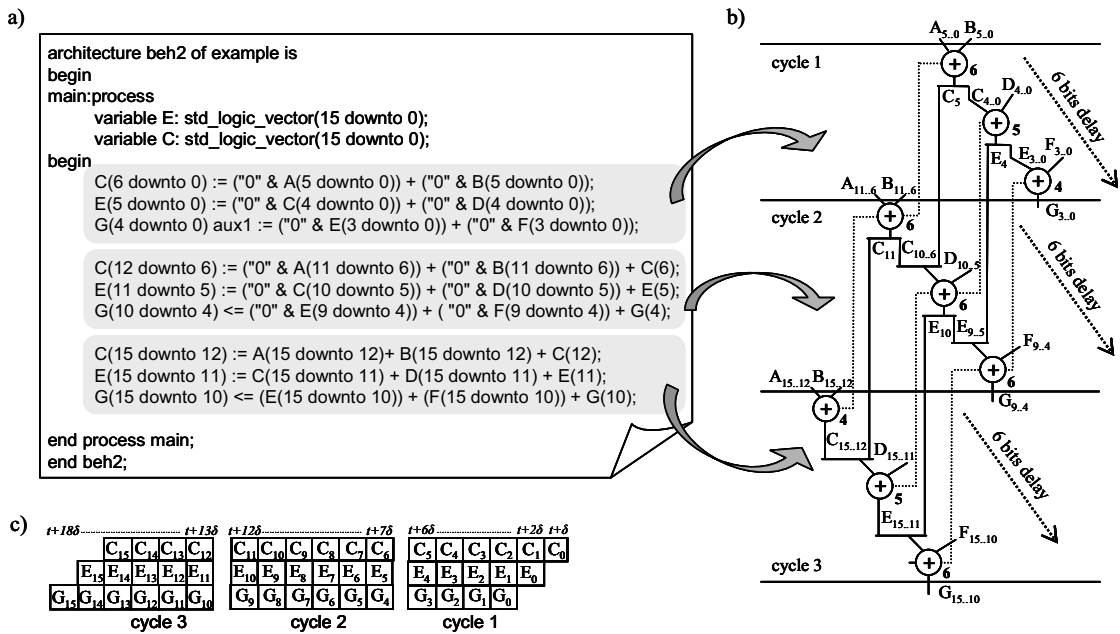

**Fig. 2.** a) Optimized specification, b) schedule of the transformed specification, c) inherent parallelism of the operations that calculate *C*, *E*, and *G*.

Table I summarizes the main features of the three implementations. The values shown have been produced by Synopsys Design Compiler after logic synthesis, and include, in all cases, the routing and controller costs. The execution time of the implementation synthesized from the transformed specification is comparable to that obtained using chaining techniques. However, the area is quite smaller if the optimized specification is used. Note that it is even smaller than the area of the circuit obtained from the schedule shown in Fig. 1 b).

In the example the additions of the specification are executed over ripple-carry adders. Nevertheless big reductions in both the cycle length and the datapath area can also be achieved by using faster and more expensive adders (carry-lookahead, fast lookahead, and carry-save).

**Table I. Comparison of the implementations in Figs. 1 and 2**

|  | Original specification | | Optimized specification |
|---|---|---|---|
|  | Fig. 1 b) | Fig. 1 d) |  |
| Latency | 3 | 1 | 3 |
| Cycle length | 9.4 ns | 9.57 ns | 3.55 ns |
| Execution time | 28.22 ns | 9.57 ns | 10.66 ns |
| FU cost | ⊕ 16 bits (162 gates) | 3 ⊕ 16 bits (486 gates) | 3 ⊕ 6 bits (176 gates) |
| Registers cost | ⊕ 16 bits (81 gates) | — | 5 ⊕ 1 bit (55 gates) |
| Routing area | 2 mux 3 to 1 - 16 bits 1 mux 2 to 1 - 16 bits (176 gates) | — | 6 mux 3 to 1 - 6 bits 5 mux 2 to 1 - 1 bit (159 gates) |
| Controller area | 60 gates | 32 gates | 62 gates |
| Total area | 479 gates | 518 gates | 452 gates |

## 3. Optimization method

The optimization method improves the results obtained by HLS algorithms when solving the time-constrained scheduling problem of data-intensive applications. It transforms the behavioural specification into another one whose synthesis results in smaller execution times. Performance results are comparable to those reported by BLC techniques, but implementation areas are smaller.

During this process some operations are broken up into several smaller ones, allowing their schedule in different cycles (possibly unconsecutive). Hence, the transformed description may have more operations, and also their types

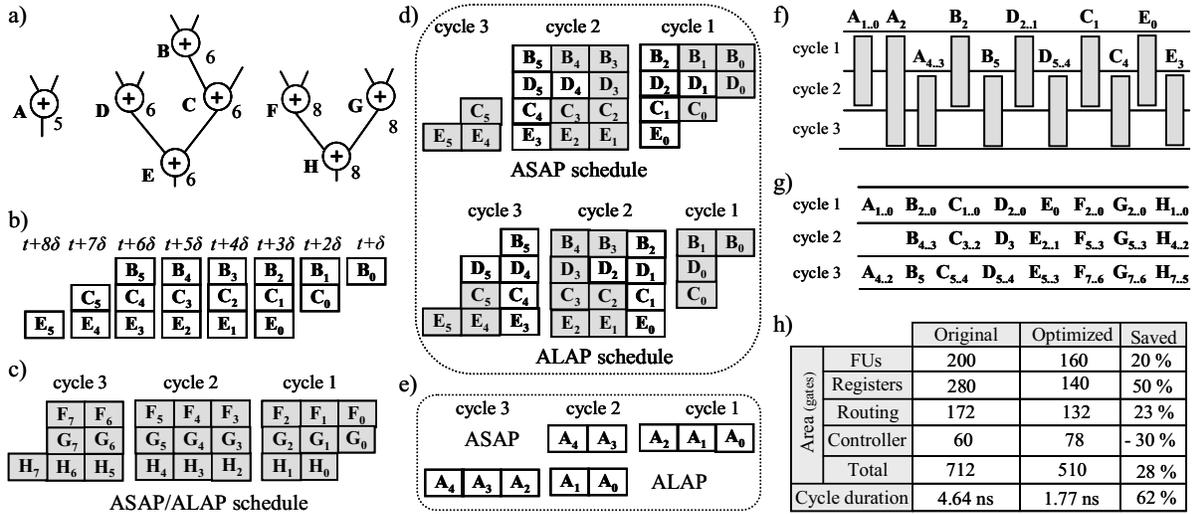

**Fig. 3.** a) DFG of one behavioural description, b) inherent parallelism of the path formed by operations *A*, *C*, and *E*, c), d) and e) ASAP and ALAP schedules of the DFG operations, f) mobilities of the unscheduled operations, g) schedule obtained from the optimized specification, and h) area and performance comparison.

and widths may be different. The set of operations to be broken up and the fragment sizes are selected considering the circuit latency and the execution times, mobilities, and data dependencies of the operations

In the present version of the algorithm we have considered signed and unsigned *additive* operations i.e. that can be transformed into additions: multiplications, subtractions, comparisons, maximum, etc. The algorithm comprises the following three phases:

1) *Operative kernel extraction.* Signed operations are transformed into several unsigned ones, and *additive* operations into additions and some glue logic.

2) *Estimation of clock cycle duration.* The critical path is identified and its length used to estimate the clock cycle.

3) *Fragmentation of operations.* Some of the operations in the behavioural description are broken up in order to fulfil the time constraint imposed in the previous phase.

### 3.1. Operative kernel extraction

In order to increase the number of operations that may share one FU, our algorithm unifies the different representation formats used in the specification. It transforms signed operations into several unsigned ones, e.g. a two's complement signed multiplication of m×n bits is transformed using our variant of the Baugh & Wooley algorithm into one multiplication of ($m$-1)×($n$-1) bits, and two additions of $m$ and $n$+1 bits. Afterwards it extracts the additive kernel of the specification operations, to be transformed into several additions and glue logic. Some of the transformations performed are shown in [8].

### 3.2. Estimation of the clock cycle duration

The critical path identification becomes the first task to be done. The critical path of a behavioural description is the path of the DFG taking the longest time to be executed. It has been measured in number of 1-bit chained additions, so non-additive operations are not considered.

To calculate the time consumed by one path, operations are crossed from its output to the input. For each operation crossed, 1 is added to the width of the last operation (the one which produces the path output). If the operation considered is wider than its successor, the number of least significant bits (LSB) truncated is also added. The algorithm below is used to compute the execution time of every path.

*time*: path execution time  *n*: number of path operations
*width(ope)*: width of *ope*  *path[i]*: *i*-th operation of the path
*truncated_rigth(ope)*: number LSB bits truncated from the result of *ope*

```
time = width(path[n]);     i = n-1;
while ( i > 0) do
    if width (path[i]) ≤ width(path[i+1])
        then time = time + 1;
        else time = time + 1 + truncated_rigth(path[i]);
    end if;
    i = i-1;
end while;
```

Once the critical path is selected, its execution time is used to estimate the cycle duration (measured in number of 1-bit chained additions). It also depends on the latency (λ).

$$cycle\_duration = \left\lceil \frac{execution\_time(critical\_path)}{\lambda} \right\rceil$$

Fig. 3 a) illustrates the DFG of a behavioural description with 4 additions of 6 bits, 3 additions of 8 bits, and 1 addition of 5 bits. The inherent parallelism of operations *B*, *C*, and *E* is shown in Fig. 3 b). Bits *i*, *i-1*, and *i-2* of operations *B*, *C*, and *E* respectively, may be calculated in parallel. The execution time of the path formed by these operations is $8 \cdot \delta$. A conventional algorithm would select this path as the critical one, because any other path has fewer operations. However, the rippling effect makes the critical path become operations *F* and *H*, and *G* and *H*, whose execution times equal *9·δ*. In order to schedule the proposed DFG in 3 cycles, the cycle duration estimated by the algorithm comes to *3·δ* (3 chained 1-bit additions).

### 3.3. Fragmentation of operations

The clock cycle duration estimated in the previous phase may be smaller than the execution time of some specification operations. In order to meet the time constraint imposed, some operations must be broken up to allow their execution in several cycles. To identify which ones must be broken up, and the number and widths of the fragments to be obtained, the ASAP and ALAP schedules of every operation bit are performed. Both schedules are calculated taking into account the maximum number of chained bits allowed in one clock cycle. If the ASAP and ALAP schedules of one operation bit coincide, then that operation bit must be executed in the cycle fixed by both schedules. Operations with some bits scheduled in different cycles must be broken up. Additionally, operations whose bits have different ASAP and ALAP schedules are also broken up to avoid any reduction in their mobilities. The number of fragments obtained from one operation equals the number of different (ASAP schedule, ALAP schedule) pairs found in the calculation of every operation bit mobility. And the width of every fragment is the number of operation bits with the same ASAP and ALAP schedules. As a result all the fragments of the same original operation have different mobilities. The algorithm below is used to calculate the number and width of the fragments obtained from every operation.

*n_bits*: number of chained addition bits allowed in every cycle
*ASAP(ope)*: first cycle where it is possible to schedule operation *ope*
*ALAP(ope)*: last cycle where it is possible to schedule operation *ope*
*sched_ASAP[ope,i]/sched_ALAP[ope,i]*: maximum number of bits of operation *ope* that can be scheduled in cycle *i*
*fragments[ope,k].(size, ASAP, ALAP)*: set of fragments (from 0 to *k*-1) obtained from *ope*, of width *size* and mobility ASAP-ALAP cycles

```
w = width(ope);   i = ASAP(ope);   j = ALAP(ope);
while ( w > 0) do
      if  (w > n_bits) then
             sched_ASAP[ope,i] = n_bits;  sched_AlAP[ope,j] = n_bits;
      else  sched_ASAP[ope,i] = w;       sched_ALAP[ope,j] = w;
      end if
      w = w – n_bits;  i = i+ 1;  j = j - 1;
end while;
i = ASAP(ope);   j= ASAP(ope);   k = 0;
while (i ≠ ALAP(ope)) and (j ≠ ALAP(ope)) do
      while( sched_ASAP[ope,i] = 0) do i = i+1 endwhile;
      while( sched_ALAP[ope,j] = 0) do j = j+1 endwhile;
      M = Min(sched_ASAP[ope,i], sched_ALAP[ope,j]);
      sched_ASAP[ope,i] = sched_ASAP[ope,i] –M;
      sched_ALAP[ope,j] = sched_ALAP[ope,j] –M;
      fragments[ope, k].size = M;    fragments[ope, k].ASAP = i;
      fragments[ope, k].ALAP = j;    k = k + 1;
end while;
```

These fragmentations produce new data dependencies among operations and operation fragments. The execution of one fragment requires the previous execution of the precedent LSB of the same operation (to use the carry out produced as its carry in), and also the bits used as input operands. These new data dependencies and the mobilities of operations and fragments just calculated must be taken into account during the scheduling.

Figs. 3 c), d) and e) show the ASAP and ALAP schedules of the operations in the example. They have been calculated taking into account the cycle duration constraint computed previously (3 chained 1-bit additions). Both ASAP and ALAP schedules coincide on operations *F*, *G*, and *H*. This means that their mobilities include just one cycle, and in consequence they are already scheduled, as depicted in Fig. 3 c). In the schedule proposed, operation *F* is fragmented into $F_{2..0}$, $F_{5..3}$, and $F_{7..6}$, in cycles 1, 2, and 3 respectively. The ASAP and ALAP schedules differ on several bits in the remaining operations, as Figs. 3 d) and e) illustrate. In order to avoid reductions in their mobilities, these operations must be broken up. For example, operation *B* is broken up into $B_{1..0}$, $B_2$, $B_{4..3}$, and $B_5$. Both the ASAP and ALAP schedules of fragments $B_{1..0}$ and $B_{4..3}$ coincide, therefore they are already scheduled in cycles 1 and 2 respectively. Other fragments of operation *B* are not scheduled yet. The mobility of $B_2$ includes cycles 1 and 2, and the mobility of $B_5$ cycles 2 and 3. Note in grey color the bits already scheduled. Fig. 3 f) shows the mobility of the unscheduled fragments. The optimized specification consists of the scheduled fragments shown in Figs. 3 c) and d), and the unscheduled ones in Fig. 3 f).

The schedule obtained by a conventional algorithm from the optimized specification is shown in Fig. 3 g). In order to balance the number of operations executed per cycle, operation *A* is calculated in cycles 1 and 3. The fragmentation of operations performed by the optimization algorithm allows a conventional scheduler to produce schedules where some operations can be calculated during several unconsecutive cycles. To our knowledge, there is not any other design technique able to allow the execution of one operation in several unconsecutive cycles with the aim of improving the circuit performance. Fig. 3 h) compares the implementations synthesized from both the optimized specification and the original one, being the latency 3 cycles in both cases. In addition to the huge clock cycle reduction (62%) a substantial area saving has also been achieved (28%).

### 4. Experimental results

In order to evaluate the optimization method, we have synthesized (using Synopsys Behavioral Compiler, BC, version 2001.08) a set of specifications. For each one the Behavioral Compiler was applied on:
- the original specification
- the specification obtained after the application of the presynthesis transformations presented in this paper.

In all cases, best results have been achieved from the optimized specifications with negligible increments in the design time. The experimental work includes the optimization and subsequent synthesis of several classical HLS benchmarks [9], and part of a real application.

The classical benchmarks synthesized are a fifth order elliptical wave filter (elliptic), a differential equation solver

**Table II. Synthesis of some classical HLS benchmarks**

| | λ | Cycle duration (nanoseconds) | | | Area increment |
|---|---|---|---|---|---|
| | | Original | Optimized | Saved | |
| elliptic | 11 | 51.59 | 11.63 | 77.45 % | 5.4 % |
| elliptic | 6 | 60.45 | 21.21 | 64.9 % | 6.45 % |
| elliptic | 4 | 68.2 | 29.4 | 56.89 % | 8.23 % |
| diffeq | 6 | 94.45 | 39.85 | 57.8 % | 4.57 % |
| diffeq | 5 | 97.56 | 46 | 52.84 % | 5.98 % |
| diffeq | 4 | 101.34 | 59.03 | 41.75 % | 9.04 % |
| iir4 | 6 | 93.6 | 15.28 | 83.67 % | 5.76 % |
| iir4 | 5 | 93.6 | 18.41 | 80.33 % | 7.34 % |
| fir2 | 5 | 94.57 | 14.5 | 84.67 % | 6.03% |
| fir2 | 3 | 94.57 | 20.8 | 78 % | 6.78% |

(diffeq), a fourth order IIR filter (iir4), and a second order FIR filter (fir2). Table II shows the clock cycle duration and the datapath area comparison between the implementations obtained from the transformed specification and from the original one for several different latencies (λ). Performance has been improved 67% on average, and reductions of the cycle length of up to 84% have been obtained. The datapath area has augmented 6% on average. The number of operations in the transformed specification is around 34% larger on average.

We have also synthesized part of a real circuit description, the ADPCM decoding algorithm specified in the Recommendation G.721 of CCITT. The modules synthesized are: Inverse *Adaptive Quantizer* (IAQ), *Tone & Transition Detector* (TTD), *Output PCM Format Conversion* (OPFC), and *Synchronous Coding Adjustment* (SCA). OPFC and SCA modules have been synthesized together, and IAQ and TTD independently. The latencies used to synthesize the original and the optimized specifications are the ones selected by BC in the conventional schedule (using the command *schedule – io_mode free_floating*). Table III shows the cycle length of the schedules obtained from both specifications. The circuit performance has been improved 66% on average. Additionally the circuit area has been reduced 4% on average, thanks mainly to the normalization of types and formats performed during the operative kernel extraction phase. The number of operations in the optimized specification has augmented around 30%.

In all the experiments performed the cycle length saved has grown with the circuit latency. To illustrate this dependency we have scheduled a behavioural description using both the original and the optimized specifications for different values of the circuit latency. Fig. 4 shows graphically how the curves (the cycle length of the schedules obtained from both specifications) diverge as the latency becomes bigger.

**Table III. Synthesis of some modules of ADPCM decoder**

| Module | λ | Cycle duration (nanoseconds) | | | Area saved |
|---|---|---|---|---|---|
| | | Original | Optimized | Saved | |
| IAQ | 3 | 6.96 | 2.4 | 65.51 % | 2.4 % |
| TTD | 5 | 9.28 | 3.66 | 60.56 % | 6.25 % |
| OPFC + SCA | 12 | 9.39 | 2.36 | 74.86 % | 3.26 % |

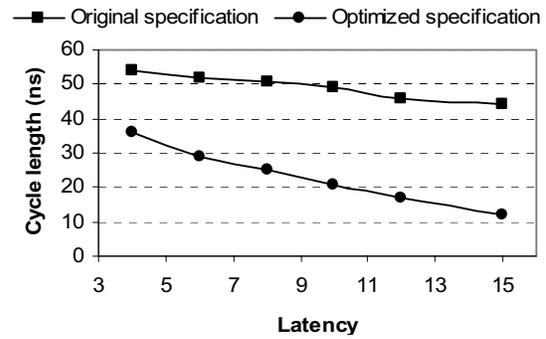

**Fig. 4. Cycle length of the schedules obtained from the original and optimized specifications.**

## 5. Conclusion

This paper presents a presynthesis optimization process that transforms a behavioural specification into a new one, whose schedule results in huge improvements of circuit performance. The specification transformation performed is based on an estimation of the clock cycle duration, used to select the operations to be broken up and the number and widths of the fragments to be obtained. These fragmentations allow a conventional scheduler to select a set of possibly unconsecutive cycles to execute one operation, by assigning separately its fragments to different cycles in the new specification. Additionally, the result bits of every operation executed in several cycles are available the cycle they are calculated in, to be used by any successor. Experimental results show reductions of up to 85% on the cycle duration in the circuits synthesized.